\documentclass[9pt]{article}
\usepackage[utf8]{inputenc}

\usepackage{amsmath}
\usepackage{graphicx}

\usepackage{amssymb}
\usepackage{authblk}

\author[*,1]{Stefan Rothe}
\author[1]{Qian Zhang}
\author[1]{Nektarios Koukourakis}
\author[1]{Jürgen W. Czarske}

\affil[1]{Technische Universität Dresden, Faculty of Electrical and Computer Engineering, Laboratory of Measurement and Sensor System Technique, 01062 Dresden, Germany}
\affil[*]{Corresponding author: stefan.rothe@tu-dresden.de}
\date{}                     
\title{Intensity-only Mode Decomposition on Multimode Fibers using a Densely Connected Convolutional Network}

\usepackage[dvipsnames]{xcolor}
\usepackage{graphicx}
\usepackage{xfrac}
\usepackage{amsmath}
\usepackage{siunitx} 
\usepackage{makecell}
\usepackage{hyperref} 
\hypersetup{linkbordercolor=green}
\begin{document}

\maketitle

\section{Abstract}
The use of multimode fibers offers advantages in the field of communication technology in terms of transferable information density and information security. For applications using physical layer security or mode division multiplexing, the complex transmission matrix must be known. To measure the transmission matrix, the individual modes of the multimode fiber are excited sequentially at the input and a mode decomposition is performed at the output. Mode decomposition is usually performed using digital holography, which requires the provision of a reference wave and leads to high efforts. To overcome these drawbacks, a neural network is proposed, which performs mode decomposition with intensity-only camera recordings of the multimode fiber facet. Due to the high computational complexity of the problem, this approach was usually limited to a number of 6 modes. In this work, it is shown for the first time that by using a DenseNet with 121 layers it is possible to break through the hurdle of 6 modes. The advancement is demonstrated by a mode decomposition with 10 modes experimentally. The training process is based on synthetic data. The proposed method is quantitatively compared to the conventional approach with digital holography. In addition, it is shown that the network can perform mode decomposition on a 55-mode fiber, which also supports modes unknown to the neural network. The smart detection using a DenseNet opens new ways for the application of multimode fibers in optical communication networks for physical layer security. 

\section{Introduction}
the use of optical multimode fibers (MMF) is regarded as a key-element in the evolution of optical communication networks \cite{richardson2013}. Under the use of MMFs, the transferable data density is significantly increased compared to singlemode fiber (SMF) based optical networks. The difference is that an MMF enables mode division multiplexing (MDM) \cite{ryf2011, ryf2018}, whereas an SMF provides only one mode for spatial transmission. In addition, MMFs offer an enhancement in information security, as during controlled data transmission the complex distortion of the light field induced by the MMF can be exploited in such a way that a potential eavesdropper is confronted with serious challenges to decode the sent message. This technique is called physical layer security and was first introduced by the authors using an MMF \cite{rothe2020}.\\
Due to its complex light propagation property a robust and practical calibration of the waveguide is required to enable commercial use of MMFs in communications engineering. The transfer function of the fiber must be known for controlled data transmission between input and output
. One possibility is the measurement of the transmission matrix (T) of the MMF. There are several ways to determine T of an MMF. Recently, the first experimental research on T determination with only one active MMF access was presented \cite{chen2020}. In this case, so-called spatial pilots are used as passive receiver coding. A tunable laser is used to obtain the necessary phase information by scanning the frequency domain and thus the entire channel can be estimated. This approach is very promising, especially because only one optical access is used and a transmitter-side MIMO can be implemented. However, the channel estimation requires a lot of computational power, especially with the use of a high mode number, which is why this approach will require a comprehensive hardware implementation. \\ 
By using both optical accesses of the MMF, a fast calibration of the MMF can be achieved. Digital Optical Phase Conjugation (DOPC) is a commonly used method to enable light control through an MMF \cite{buttner2020velocity, czarske2016transmission} or coherent fiber bundle \cite{kuschmierz2018self}. However, the application of DOPC is limited to the reconstruction or superposition of guide stars and is not used to generate arbitrary output signals. This is made possible by measuring the complete T, even in real time \cite{caravaca2013}. For this purpose, it is important to decide in which domain the MMF will be used. As a thin endoscope for image transmission, it can be useful to interpret the transmitted images as superposition of diffraction limited foci \cite{ploschner2015}. In fields of communication technology where MMF is used for MDM, it is advantageous to use T in the basis of the modes supported. In this case, a Spatial Light Modulator (SLM) is used to sequentially excite the individual modes at the input of the MMF. At the output of the MMF, the transmitted light field must be decomposed into its individual components, which is called mode decomposition (MD). This is done by taking advantage of the fact that an arbitrary field distribution on the MMF facet $\underline{E}_{\text{MMF}}$ is a complex superposition of the individual mode fields supported by the MMF $\underline{E}_{\text{mode},i}$. Each mode has its own mode weight, consisting of amplitude $\rho$ and phase $\phi$, which corresponds to the respective complex part of the light field distribution of the MMF: 
\begin{equation}
    \underline{E}_{\text{MMF}}=\sum_i^N{\rho_{i}e^{\text{j}\phi} \underline{E}_{\text{mode},i}} \text{ , }
\end{equation}
where N indicates the number of modes the MMF supports. Holographic \cite{rothe2019,mazur2019characterization} or scanning \cite{carpenter2012,carpenter2014,fontaine2019laguerre} approaches have been presented for enabling the mode decomposition experimentally. In contrast to the holographic approach, the scanning approach works without a reference and requires $N^2$ measurements to obtain the complete T. The holographic approach requires only $N$ measurements for the complete T, which is an advantage particularly in highly dynamic communication applications. However, holographic MD is impractical for use in a communications network, as each MMF used for data transmission requires a second SMF for delivery of the reference beam for the holographic phase measurement. Intensity-only approaches are thus beneficial, since they do not depend on a phase measurement. This is why several numerical approaches have been introduced, which require a less complex optical setup. Different optimization methods can be used to perform a intensity-only MD. However, they depend very much on the initial values of the mode weights to be determined \cite{huang2015,li2017,an2019numerical}. If the initial states are chosen unfavourably, a local optimum which differs from the global optimum can be found. This is not the case for iterative methods like the Gerchberg-Saxton algorithm \cite{bruning2013}, which is, however, time-consuming due to the iterative approximation.\\
A computer-aided approach, which in comparison to iterative or optimizing approaches does not consider each problem individually, is the use of a neural network \cite{barbastathis2019use}. By means of a training procedure the weights and biases of the network are adjusted a priori to the application. First alternative approaches deal with the circumvention of this often very time-consuming procedure by connecting an untrained neural network with a physical model to perform the desired tasks \cite{wang2020phase}. Here, the required weights are found via an optimization problem. However, the time required to reconstruct a recording will increase exponentially, which means that the real-time capability of the system is no longer guaranteed. In this manuscript, a classical training procedure is used for adjusting the weights and biases before application. Once these have been set, the network can make predictions without being retrained.\\
In various applications with MMFs, neural networks are becoming more and more popular to solve various imaging problems. For example, Convolutional Neural Networks (CNN) have been used to reconstruct an image that is scattered by the complex transmission characteristics of an MMF \cite{fan2019deep,borhani2018learning}. For the MD of an MMF, which is the subject of the work presented, a CNN is also used. It has already been shown that complex problems using purely intensity-based signal evaluation employing neural networks provide a yield comparable to that of a holographic counterpart \cite{kakkava2019,rothe2020deep}. For the present problem according a MD, the goal is to transmit a pure intensity image into the CNN, which predicts the complex mode weights. An et al. have introduced a way to discriminate between 3 or 6 modes of few-mode fibers \cite{an2019} using a CNN-type neural network. For this purpose a Very Deep Convolutional Network (VGG), which consists of 16 layers, was trained and tested \cite{simonyan2014}. The predicted vector of complex mode weights is used to reconstruct intensity distribution, which is then compared with the measured image by means of a cross correlation. However, due to the rapidly increasing complexity for each additional mode to be discriminated, a limit of 6 distinguishable modes is reached. So far, it could not be shown that more than 6 modes can be distinguished with the VGG proposed. It is expected that for a higher number of modes the training data or the network used must be modified. \\
 To overcome this limitation, a CNN-based approach to distinguish between $N=10$ modes is presented in this manuscript by using a Densely Connected Convolutional Network (DenseNet) \cite{huang2017}. The DenseNet presented is trained in a procedure with GPU parallelization and transfer learning, which will be explained in the following. The CNN is only trained with simulation data offline and runs in an experimental environment with a 10-mode MMF, where an MD is performed. In addition, using a 55-mode MMF, MDM at the 10 modes known is performed by employing inverse precoding. It can be shown, that the CNN-based MD works properly within this scenario. 
 The approach presented is compared to conventional holographic MD and paves the way for the practical implementation of MMF based optical networks using MDM and physical layer security.
 
\noindent The manuscript is structured as follows: first a basic introduction to DenseNet and the generation of appropriate training data for the MD based on CNNs is given. Then the training procedure with multi-GPUs and transfer learning, which was especially developed for DenseNet, is explained before a validation of the trained network is performed. Finally, the optical setup is presented, which is used to record experimental data. Results on intensity-based MD of random MMF signals at 10 modes are presented and discussed. Finally, it is shown that when implementing MDM on a 55-mode MMF, the 10 known modes can be detected by the DenseNet. 

\section{Procedure}
\subsection{Implementation of a DenseNet for Mode Decomposition}
\begin{figure*}[tb]
\centering
(a)\includegraphics[width=1\textwidth]{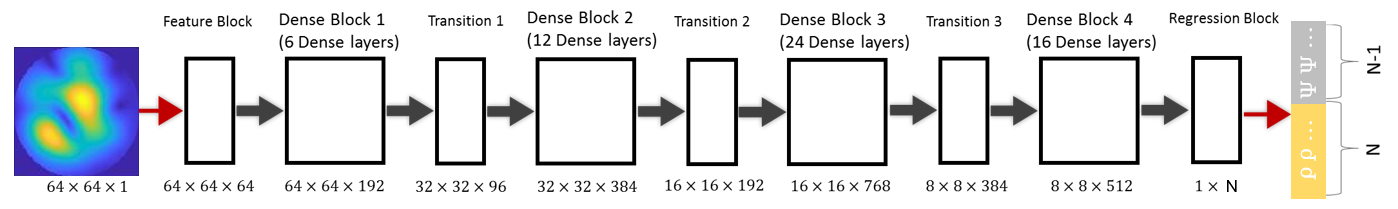}\\
(b)\includegraphics[width=0.98\textwidth]{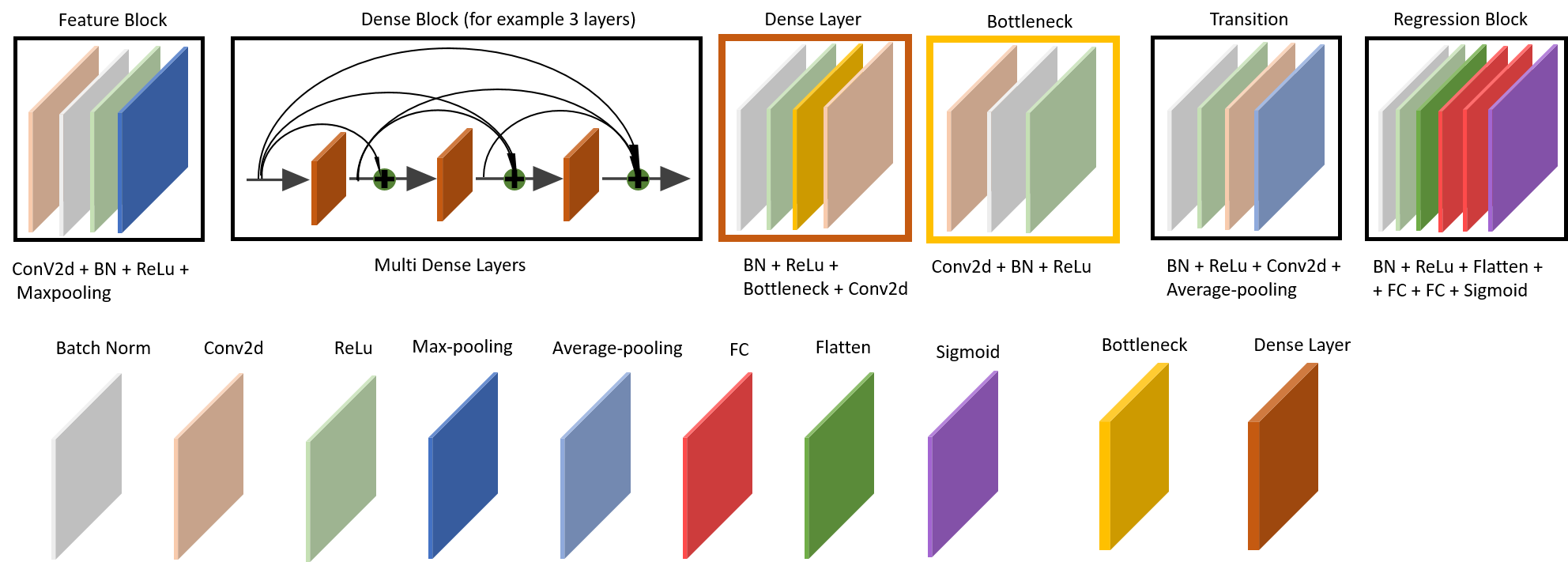}
\caption{(a) A DenseNet-type CNN with 4 Dense blocks is implemented for complex MD. One image with the size of $64\times64$ pixels is taken as the input of the network. The output is a vector with $N$ amplitude and $N-1$ phase values corresponding the the individual mode weights of the modes being decomposed. Each rectangular box represents a combination of multi-layer. The output size of feature map of each block are denoted below the boxes. (b) Detailed presentation of all hidden layers used.}
\label{fig:CNN}
\end{figure*}

The aim of the presented investigations is to train a CNN, which is able to perform a complex MD on the light field of a MMF based on simple intensity-only camera pictures. For this purpose, the DenseNet topology developed by Huang et al. is used as a regression CNN in this work \cite{huang2017}. The CNN used has a total of 121 layers and $22,963,413$ learnable parameters. In Fig. \ref{fig:CNN} its topology is visualized. 
The major characteristic of such kind of CNN is the use of Dense Blocks. Since information or features about the input of the network passes through many layers, individual details can disappear on the way to the output. The key to solve this problem is to create connections between early layers and later layers. Each layer obtains additional inputs from all previous layers and then passes its own output to all subsequent layers, that ensures maximum information flow between each layer. This is a way to consider as many details of the input as possible bundled for the evaluation at the output. In Fig. \ref{fig:CNN}(b) the layout of one Dense Block is illustrated schematically. If the amount of dense layers inside a Dense Block is $L$, the $l^{th}$ layer has $l$ inputs and its own feature maps are processed on to all $L-1$ subsequent layers. Thus, there are $\frac{L(L+1)}{2}$ connections inside an L-layer Dense Block. This design exploits a high potential of the CNN by means of feature reuse, and makes the network easier to train.

\subsection{Generation of Training Data}
For the presented problem, one pair of the training data includes a 2D intensity distribution and a label vector with the corresponding mode weights, including amplitude $\rho$ and phase $\phi$. Each change in the elements of the mode weight also results in a change in the intensity distribution. To ensure that the CNN can extract the corresponding mode weights from a given intensity image, a number of proper training data has to be at hand. Simulated intensity images serve as ground truth and are compared quantitatively with the images resulting from the network predictions during the training process. Later on, the MD is performed on experimental images. The images are assumed to be accurate enough to serve as ground truth for these cases \cite{barbastathis2019use}.

Two different approaches were used to create a synthetic database. In order to determine individual mode weights from arbitrary MMF light fields, the whole mode domain with specified mode combinations (SMC) can be scanned by two given step sizes $s_{Amp}$ and $s_{phase}$ to cover each possible case of mode combination. This approach is  promising, since the training data generated is extremely representative. However, the amount of training data $M$ is inflated rapidly with each added mode:
\begin{equation}
    M\approx \left( \frac{1}{s_{Amp}+1}\right)^N \cdot \left( \frac{1}{s_{Phase}+1}\right)^{N-1}
\end{equation}
For this reason, the SMC approach is only used for the possible extreme cases (e.g. with Mode Division Multiplexing only one mode will reach the receiver). However, in order to cover the representation of the problem as comprehensively as possible with the least amount of data, the major part of the data used consists of pseudo-random mode combinations (PRMC). Thus, the amount of data as well as the representation of the investigated problem can be controlled by using both SMC and PRMC data. All training data generated is based on a simulation and does not include any knowledge about the experimental environment. The number of data for each training step executed can be found in Table \ref{tab:cnn_training_parameter}.\\
According to the 2 methods mentioned above, the intensity-only image can be directly generated and saved, however, the label of one ensemble of the training data needs to be further processed due to phase ambiguities \cite{an2019}. The N amplitude value of the output vectors can be directly used as amplitude information $\rho$ regarding the individual mode weights. However, there are 2 undesired hurdles involved in obtaining phase information. The first one is that ensembles of mode combinations with different phase distributions can have the same amplitude distributions and thus the same intensity image. This is the case if there are different absolute phase values, but the same relative phase difference between the individual modes. Therefore, if the phase value is directly put into the label vector, the network cannot reach convergence because one intensity-only image might have several different label vectors. For this reason, the phase of the fundamental mode $LP_{01}$ is set to zero and only the relative phase difference between the higher order modes and $LP_{01}$ is taken into account. This leads to a label vector containing $N-1$ phase terms $\phi$ and solves the problem of ambiguity. The second hurdle is, that a complex-valued mode field has the same intensity-only image as its conjugated field, which is also critical for the CNN performance. For this reason, not the actual phase differences are used, but the cosine values $\Psi$ of $\phi$:
\begin{equation}
    \Psi=\text{cos}(\phi).
\end{equation}
These precautions must also be taken into account when the predicted label vectors of the CNN are evaluated. The amplitude values $\hat{\rho}$ can be obtained directly from the predicted vectors. However, the predicted phase values $\hat{\Psi}$ are firstly rescaled from $[0, 1]$ to $[-1, 1]$ and then the relative phase difference can be calculated through the arccos function. The use of arccos function leads to an ambiguity because the output value is in a range of $[0,\pi]$, while the real relative phase is in the range of $[0, 2\pi]$. This means that any evaluation of the phase via the arccos function could have two possible phase values, with positive and negative signs. In order to obtain the correct phase information $\hat{\phi}$, each possible intensity distribution is generated with all possible combinations using positive and negative signs. The $\hat{\phi}$ is then determined by calculating the cross correlation coefficients $\Gamma$ between all possible obtained intensity distributions $I_{o}$ and the target (ground truth) distribution $I_{t}$:
\begin{equation}
    \Gamma=\frac{\begin{displaystyle}\sum\end{displaystyle}_{ROI}(I_t-\overline{I_t})(I_o-\overline{I_o})}{\sqrt{\left(\begin{displaystyle}\sum\end{displaystyle}_{ROI}(I_t-\overline{I_t})^2\right)\left(\begin{displaystyle}\sum\end{displaystyle}_{ROI}(I_o-\overline{I_o})^2\right)}},
    \label{eq:correlation}
\end{equation}
where $\overline I$ indicates the mean value of the respective intensity distribution and $ROI$ the region of interest. The phase combination corresponding to the maximum correlation coefficient is then chosen as the correct prediction. With these preparations, the uniqueness of the training data can be guaranteed.

\subsection{Training process of the CNN}
The efficient training of a CNN is a particular problem that must be solved as the size of the network increases. High training times and GPU memory are usually limiting factors for a fast training process in high-dimensional neural networks. Fortunately, the involved huge amount of matrix multiplications and similar operations can be massively parallelized and thus sped up using minibatch and GPUs, which usually have thousands of operating cores. 
In this work, the minibatch size is set to 64 or 128. Two GPUs are used, which are NVIDIA GeForce GTX TITAN X with 12,212~MB and NVIDIA GeForce RTX 2080 Ti with 11,019~MB.\\
Transfer learning is an approach, which has been introduced by Pan and Yang in 2009 \cite{pan2009}. It provides a new framework for training large-scaled deep neural networks, which have to solve a problem with high complexity. The basic idea of transfer learning is to exploit the fact, that neural networks have the ability to extract the same features (for example lines or dot) from images regarding different problems of possibly different complexity. In this work, transfer learning is used, because the problem at hand is very easy to scale. For example, a network which can distinguish between 3 modes can be taken as initial state for a network which should be able to distinguish between 5 modes etc. Further, the CNN for the MD with 10 modes was trained in 3 iterations, while the result of each iteration served as initial state for the following iteration. Thus, the final DenseNet was gradually developed to the complexity of discriminating between 10 modes.\\
In Table \ref{tab:cnn_training_parameter} the relevant parameters for the training process are summarized. The data used for the 3-mode case was divided in a ratio of 8:1:1 for training, validation and test. From the data used in the 5- and 10-mode case, 1,000 each were taken for validation and test, respectively. In all cases, the learning rate was set to $10^{-3}$ in the beginning and decreased to $10^{-5}$.
\begin{table}[ht]
\caption{Parameters used for each individual training step}
\label{tab:cnn_training_parameter}
\centering
\begin{tabular}{c|ccc}
\hline
Modes& Data & Epochs & Duration\\
\hline
3 & \makecell{49,036 SMC }& 30& 3 hours\\
\hline

5 & \makecell{160,000 PRMC }& 70& 18 hours\\
\hline
10 & \makecell{13,200 SMC and 300,000 PRMC}& \makecell{20, 50 and 10}& 43 hours\\
\hline
\end{tabular}
\end{table}

\subsection{Validation of the DenseNet approach}
\begin{figure}[!h]
\centering
\includegraphics[width=3.4in]{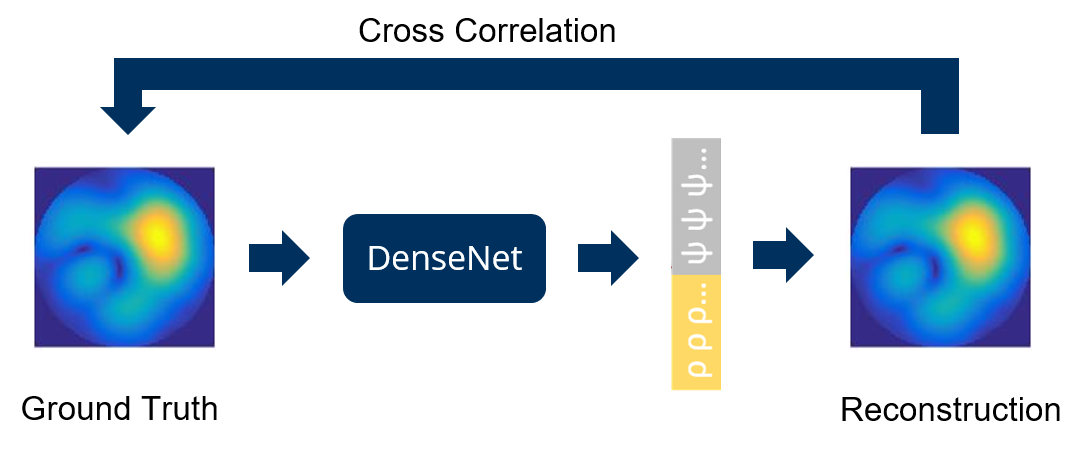}
\caption{Procedure of an CNN-based MD. Using the amplitude of an arbitrary mode field distribution, mode weights are predicted from which a mode field distribution can be calculated. The accuracy of the mode weight prediction is determined by cross correlating the calculated and the original field distribution.}
\label{fig:procedure_cnn}
\end{figure}

\begin{figure}[tb]
\centering
\includegraphics[width=3.4in]{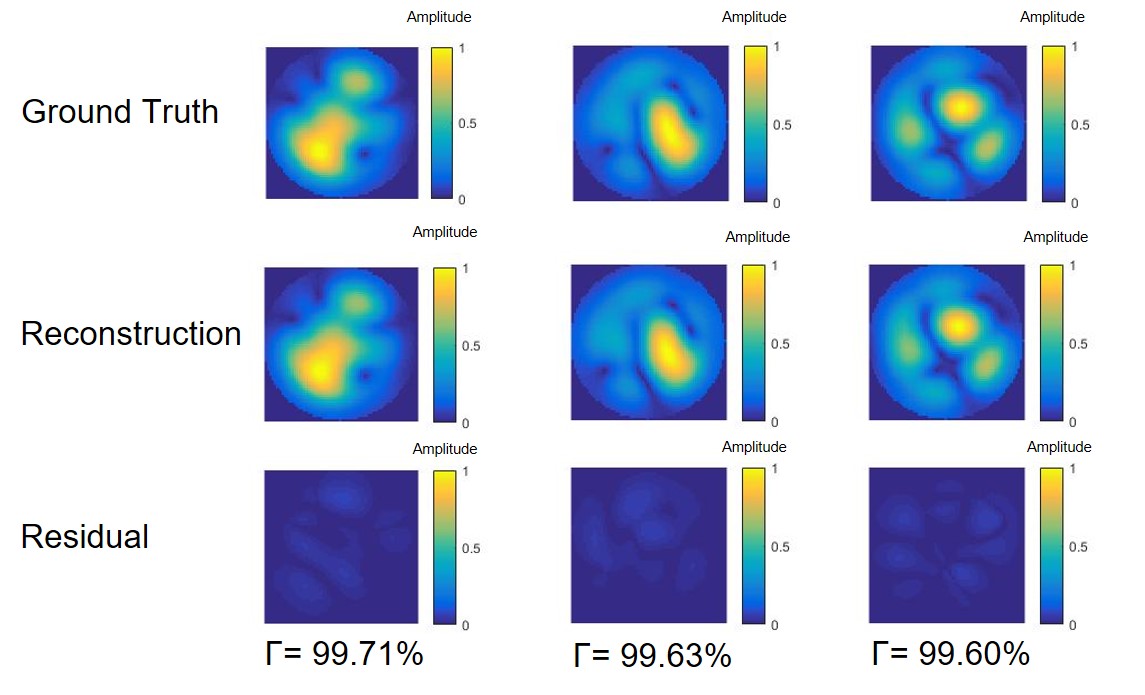}
\caption{Results of CNN based MD with 10 modes using synthetic data. The respective reconstructed amplitude field distributions, the corresponding ground truths, and the remaining residuals after subtraction are shown.}
\label{fig:ccc_synth}
\end{figure}

To evaluate the quality of the network predictions, an intensity distribution is reconstructed using the predicted mode weights, which is cross correlated with the ground truth (Eq. [\ref{eq:correlation}]). 
After each individual training process, the cross correlation value served as a quality criterion for the performance of the CNN trained.
The operation procedure of the CNN-based MD is illustrated in Fig. \ref{fig:procedure_cnn}.
In Table \ref{tab:cnn_validation}, the correlation values of the three training processes using the synthetic PRMC test data are summarized. In the 3-mode case, a correlation between reconstructed and ground truth intensity distribution of $0.9996$ can be obtained. In the 5-mode case a correlation of $0.9934$ is achieved. These values are in the same range as the comparable CNN-based MD techniques introduced by An et al. \cite{an2019}.
Further, the approach using the DenseNet-type topology also offers the possibility to distinguish between 10 modes. In this case, a correlation value of $\Gamma_{\text{CNN,synth}}=0.9762$ can be achieved. In Figure \ref{fig:ccc_synth}, three examples of reconstructed mode fields and corresponding ground truths including their residuals are shown.
\begin{table}[h]
\caption{DenseNet performance on different mode cases}
\label{tab:cnn_validation}
\centering
\begin{tabular}{c|ccc}
\hline
Number of modes & 3 & 5 & 10\\
\hline
$\Gamma_{\text{CNN,synth}}$ & $0.9996$& $0.9934$& $0.9762$\\
\hline
\end{tabular}
\end{table}

\subsection{Optical Setup}
\begin{figure}[!h]
\centering
\includegraphics[width=3.4in]{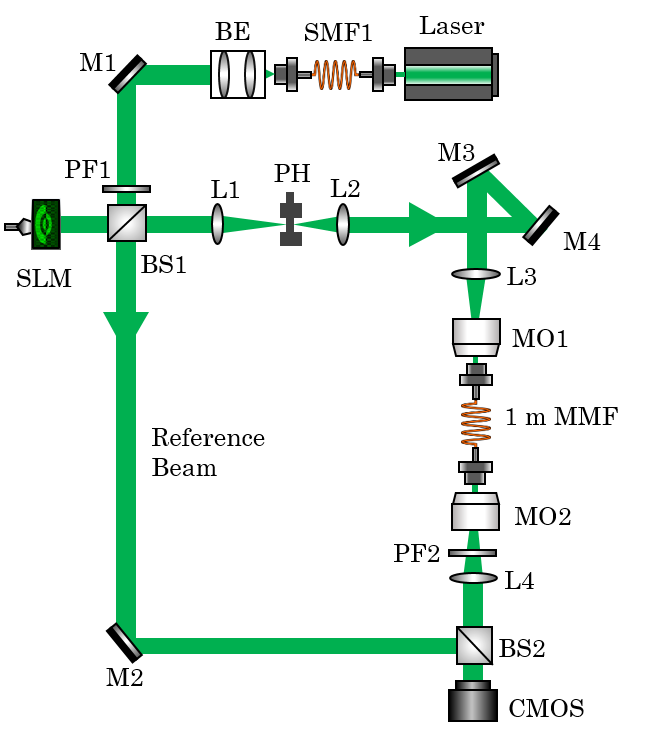}
\caption{Optical setup for the execution of the experiments. Both a holographic MD and a CNN-based MD can be performed.}
\label{fig:setup}
\end{figure}

An optical setup is developed to generate experimental data that is fed into the CNN, which was trained exclusively with synthetic data. 
Fig. \ref{fig:setup} shows the relevant part of the setup for the procedure. This setup was originally designed to measure T of an MMF, where holographic MD is performed. The introduction of holographic MD and the adjustment of the system can be found in prior work \cite{rothe2019}.
A laser with a wavelength of $\lambda=\SI{532}{\nano\meter}$ (LaserQuantum, TORUS) is used. The beam profile is filtered spatially by means of an SMF and expanded by a beam expander (BE). The light is linearly polarized by polarization optics (PF) and divided into object and reference beam. The object beam is directed to a Spatial Light Modulator (SLM, HOLOEYE Pluto), where superpixel phase masks of randomly generated mode field distributions are displayed. These phase masks are used for wavefront shaping \cite{vellekoop2007} to excite random mode field distributions at the input of the MMF. For this purpose, a pinhole (PH) and imaging system consisting of the lenses L1-L3 and microscope objective MO1 is used, which images the amplitude and phase modulated light field distribution onto the input facet of the MMF. Using mirrors M3 and M4, the incident angle of the obeject beam can be adjusted. The random light field distributions propagate through the MMF, where they arrive at the output of the MMF in a different random light field distribution caused by mode mixing effects due to light launching deviations and manufacturing tolerances. It is assumed that the statistically uniform distributed input weights of the mode domain result in the same statistically uniform distributed output weights. An imaging system consisting of the microscope objective MO2 and lens L4 maps the light field emerging from the MMF onto the CMOS camera. A polarizing filter (PF2) ensures that only one linear polarization direction is used. On the camera, the object beam interferes with the reference beam. In the investigations shown, the amplitude distribution reconstructed during the holographic MD is also used as input for the CNN ensuring a fair comparison. It is necessary to emphasize that only one direction of linear polarization is used for the investigations shown. However, it is possible to achieve polarization dependence by extending the setup, as Carpenter et al. have shown \cite{carpenter2014}. \\ 
The fibers under test are both step index MMFs at a length of $\SI{1}{\meter}$. $\text{MMF}_{10}$ has a core diameter of $\o=\SI{10}{\micro\meter}$ and a numerical aperture of NA=0.1. $\text{MMF}_{55}$ has a core diameter of $\o=\SI{25}{\micro\meter}$ and a numerical aperture of NA=0.1. At the given wavelength, $\text{MMF}_{10}$ supports 10 different spatial modes per polarization direction, whereas $\text{MMF}_{55}$ supports 55 modes per polarization direction. 

\section{Experimental Investigations}
\subsection{Mode Decomposition at a 10-mode MMF}
\begin{figure}[!h]
\centering
\includegraphics[width=3.5in]{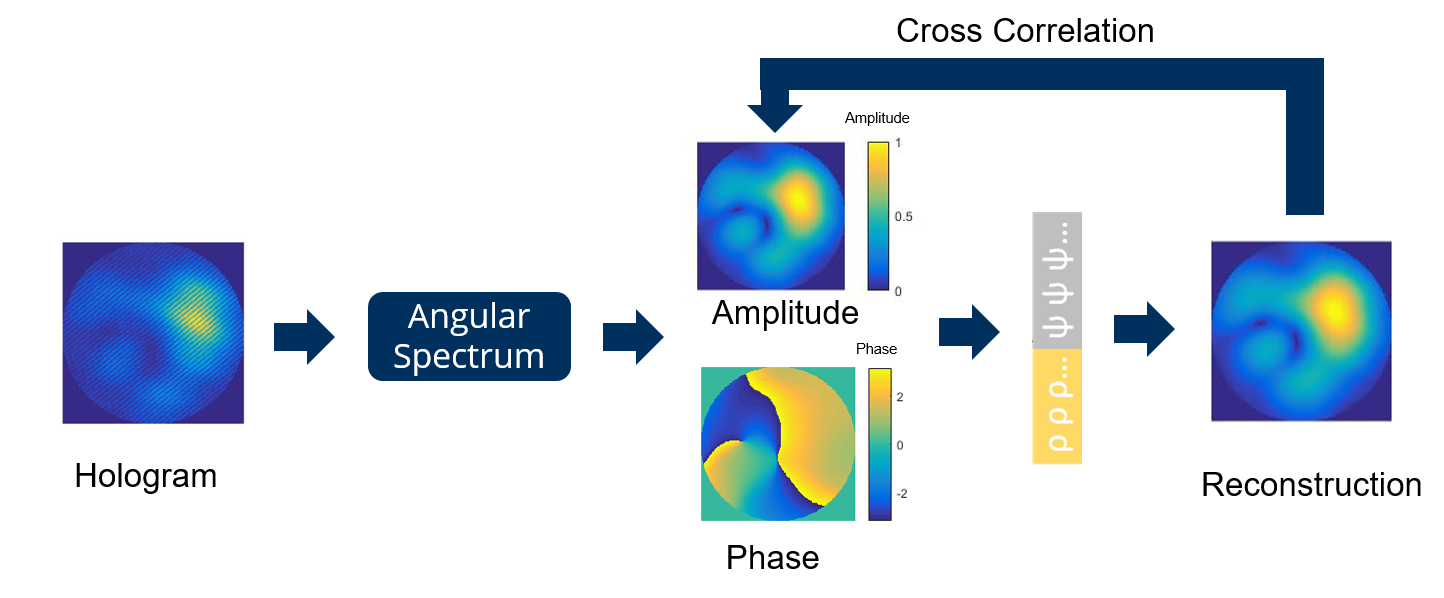}
\caption{Procedure of holographic MD. Using the Angular Spectrum method, amplitudes and phase distribution are determined from a hologram. These are put together as a complex field distribution with which a computational mode decomposition can be performed and the mode weights determined. Based on the mode weights, a mode field distribution is reconstructed whose amplitude is cross-correlated with the measured amplitude.}
\label{fig:procedure_holo}
\end{figure}

$\text{MMF}_{10}$ is integrated into the setup, first. 200 uniform distributed random combinations of mode field distributions are excited at the MMF input using an SLM. The resulting output fields are imaged on the CMOS camera, where they are recorded holographically. Amplitude and phase distributions are reconstructed by using the Angular Spectrum method \cite{koukourakis2011photorefractive}. Both amplitude and phase distributions are used for the holographic MD and thus the complex weights are calculated directly. In Fig. \ref{fig:procedure_holo} the procedure of the holographic MD is illustrated. For the CNN-based MD, only the amplitude distributions are used by employing them as input images or ground truth.

\begin{figure}[!h]
\centering
\includegraphics[width=3.4in]{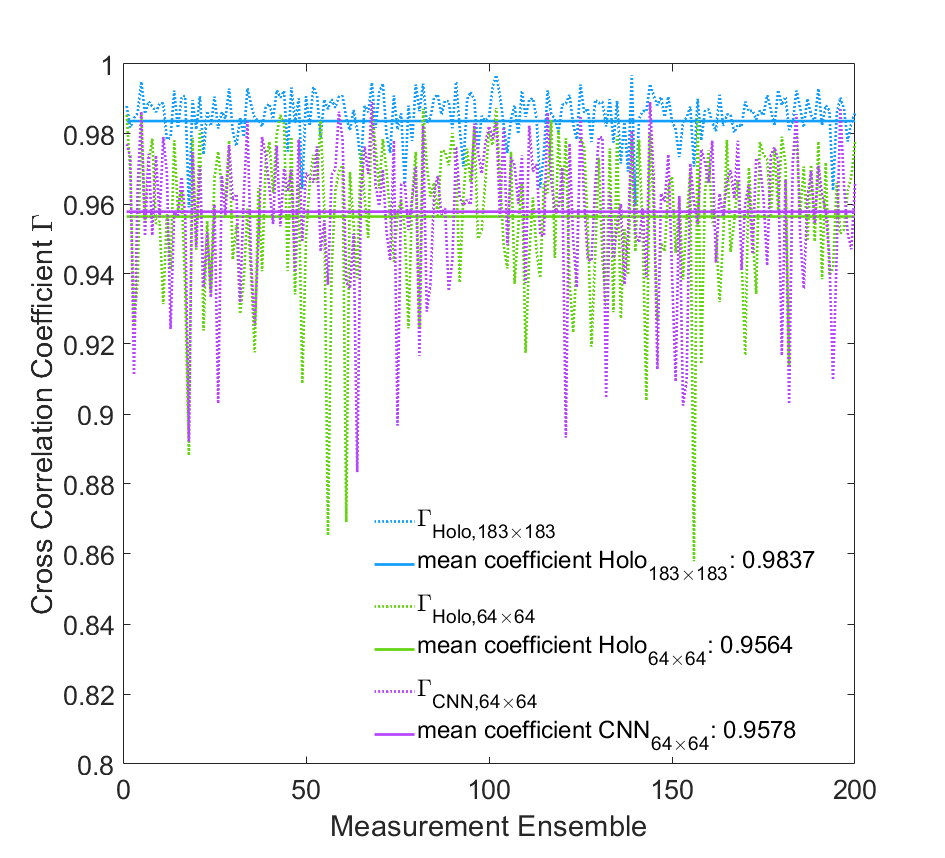}
\caption{Cross Correlation Coefficients based on holographic captured fiber output light fields using $\text{MMF}_{10}$. 200 measurement ensembles are generated}
\label{fig:ccc_holo_cnn}
\end{figure}

\noindent Using the respective mode weights, amplitude distributions are generated and cross correlation coefficients between ground truth and reconstructed distributions are determined (Eq. [\ref{eq:correlation}]). The results of the cross correlation are shown in Fig \ref{fig:ccc_holo_cnn}. A mean cross-correlation coefficient of $\Gamma_{\text{Holo,183$\times$183}}=0.9837$ could be achieved using the holographic MD, whereas the CNN-based MD yields to a mean cross-correlation coefficient of $\Gamma_{\text{CNN,64$\times$64}}=0.9578$, which is $98.12\%$ of the ideal case of ($\Gamma_{\text{CNN,synth}}=0.9762$) based on synthetic data. 
When comparing the two methods, it was ensured that one and the same amplitude distribution is used in both holographic and CNN-based MD. However, the CNN was trained with 64$\times$64 pixels, which means that the size of the recorded intensity distribution must be adjusted. This fact leads to several negative effects on the quality of the CNN-based MD. On the one hand, less degrees of freedom are taken into account in the CNN-based MD than in the holographic MD, which was recorded with the original 183$\times$183 pixels. This can lead to a worse evaluation quality. In addition, numerical artefacts occur at the edges of the measured distribution during downsampling due to random rounding errors. The problem is that the edge areas have the same portion in the determination of the cross correlation coefficients as information from the center of the image with a high signal component. This effect becomes obvious when the holographic MD is repeated with downsampled amplitude and phase distributions. Fig. \ref{fig:ccc_holo_cnn} shows the result of this investigation. The holographic MD now has a lower mean cross correlation coefficient of $\Gamma_{\text{Holo,64$\times$64}}=0.9564$ and has similar scattering behavior around the average as the CNN-based MD. The downsampling problem would be considered in a system tailored to CNN-based analysis by optimizing and fixing the image size of the MMF facet to the required number of camera pixels.

\noindent According to the studies shown, the CNN approach proves that it is suitable as a reference-free alternative for the evaluation of MMF signals. It should be remembered, that the CNN has been trained only with synthetic data and works in an experimental environment successfully. The predicted mode weights can be further used for a robust control \cite{rahmani2020actor}, which is commonly performed in imaging applications based on a field programmable gate array \cite{radner2020field,caravaca2013}. 

\subsection{Mode Decomposition for Physical Layer Security}
\begin{figure}[!h]
\centering
(a)\includegraphics[width=1.5in
]{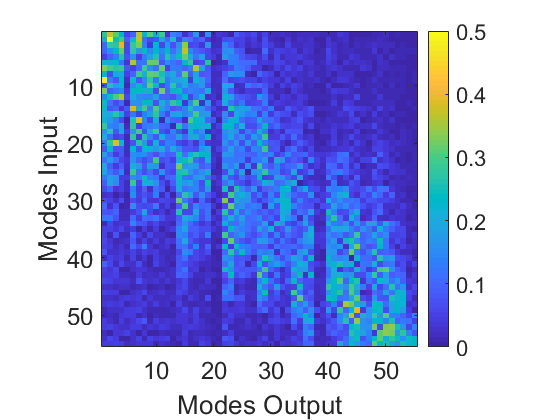}
(b)\includegraphics[width=1.5in
]{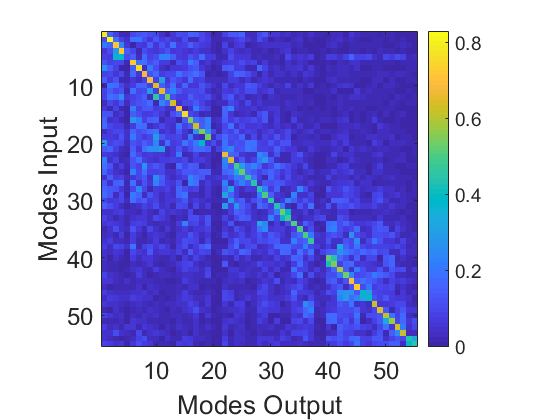}
\caption{T measurement of $\text{MMF}_{55}$ (a) before and (b) after implementing inverse precoding.}
\label{fig:T_measurement}
\end{figure}

For a practical application of Physical Layer Security in an optical network, it is necessary to use an MMF with a significantly higher number of modes than 10. The reason for this is that the transmission behaviour of the MMF becomes more complex with each additional mode, which can be exploited to the advantage of legitimate communication partners in terms of information security. A conventional implementation of physical layer security uses holographic measurement of T of an MMF to perform inverse precoding. This enables modes to be sent to a legitimate recipient of messages, while at the same time a crucial SNR disadvantage for an unauthorized recipient can be created.\\
In this work, inverse precoding was performed on $\text{MMF}_{55}$. Fig. \ref{fig:T_measurement}(a) shows the measured T. After the implementation of inverse precoding, the re-measured T has a dominant main diagonal, as shown in Fig. \ref{fig:T_measurement}(b). Modes can be generated at the output facet of $\text{MMF}_{55}$ an thus MDM is enabeld. 

\begin{figure}[!h]
\centering
\includegraphics[width=3.4in
]{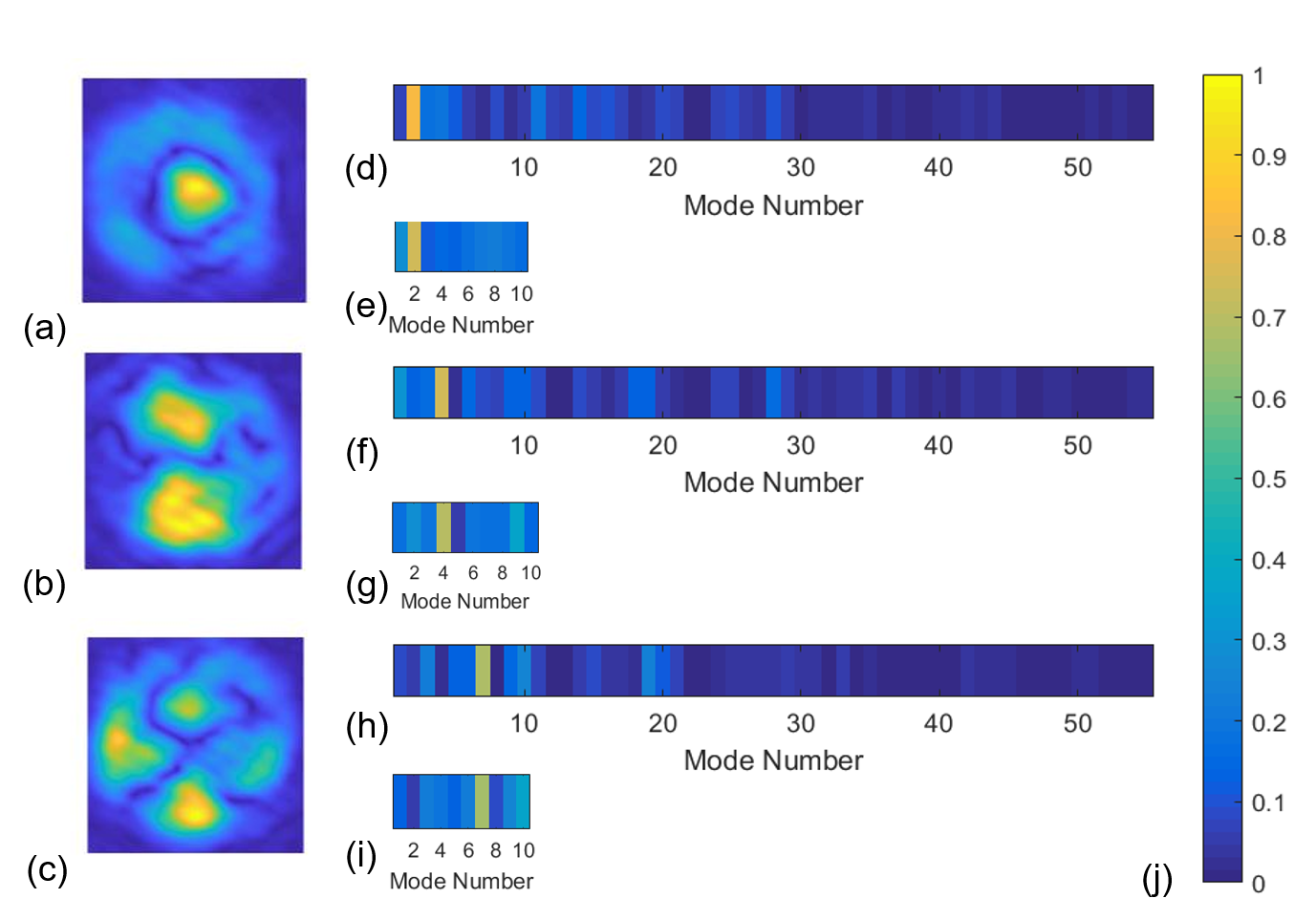}
\caption{Mode detection of three modes of $\text{MMF}_{55}$ using both CNN-based and holographic MD. Inverse precoding was performed to observe modes at the MMF output facet. (a) measured amplitude distribution of $\text{LP}_{\text{02}}$, (c) $\text{LP}_{\text{11,o}}$ and (d) $\text{LP}_{\text{21,o}}$. Using holographic MD, the observed modes can be detected. For better comparison, the first 10 elements of the MD vectors of the holographic MD are arranged in the same manner as the MD vectors of CNN-based MD. (d), (f) and (h) show the amplitude vectors of the holographic MD considering the whole 55-mode domain, whereas (e), (g) and (i) show the respective vectors of CNN-based MD. The vectors in (d) and (e) are corresponding to the MD of $\text{LP}_{\text{02}}$ shown in (a). The same applies to the vectors from (f), (g) and the image in (b), as well as to the vectors from (h) and (i) and the image shown in (c). (j) shows the colorbar used for the results presented.}
\label{fig:MDM}
\end{figure}

 \noindent Fig. \ref{fig:MDM}(a)-(c) show the mode fields of three spatial modes measured at the output facet of the fiber. The received modes can now be detected using holographic or CNN-based MD. Fig. \ref{fig:MDM}(d)-(i) show the respective MD vectors (amplitude only) of both approaches.  
 It can be seen that the respective modes can be detected, successfully. It should be remembered that both the amplitude and the phase distribution of the measured mode field are involved in the holographic MD, whereas only the amplitude information is used for the CNN-based approach. The modes shown are also included in the mode domain of $\text{MMF}_{10}$, for which the CNN was trained. If the measured mode field distributions from Fig. \ref{fig:MDM}(a)-(c) are used as input of the CNN, the modes can also be detected, although the CNN has no knowledge about the other 45 modes from  $\text{MMF}_{55}$ and was trained with synthetic data. Fig. \ref{fig:MDM}(e), (g) and (i) show the vectors from the CNN prediction. It can be seen that the bright stripes, which are representative of the detected modes, appear in the same position as in the holographic MD. This experiment clearly proves the robustness of the CNN in unknown environments. Theoretically it is  possible to train the network to other modes, which may have higher spatial frequencies. However, for this purpose it must be ensured that there are enough pixels for the spatial resolution of the degrees of freedom to train a CNN with sufficiently high performance.

\section{Conclusion}
In a specially developed training procedure, a 121 layer DenseNet-type convolutional neural network was trained successfully, which performs a complex mode decomposition at 10 modes on the basis of intensity-only camera images of multimode fiber light field distributions. The DenseNet was trained with purely synthetic data and has no knowledge of the experimental environment. Nevertheless, a mean cross correlation coefficient of 0.9578 could be achieved for experimentally recorded light fields, whereas the conventional approach using digital holography with identical data reaches a coefficient of 0.9837. The measurement results shown highlight the potential of reference-free mode decomposition based on a neural network. Furthermore, it could be shown that detection with the neural network also works with an multimode fiber that also carries unknown modes. Inverse precoding was used to perform multiplexing over the 10 known modes in a 55-mode fiber. Although the experimental recordings of the fiber signals contain degrees of freedom and features unknown to the network, the received modes can be detected. The results show no significant differences to the holographic approach. The investigations shown prove that for reference-free mode decomposition an approach based on a neural network offers a serious alternative to conventional holographic methods. These findings are of significant importance for the path to the commercial development of physical layer security in multimode fiber communication systems.

\end{document}